\documentclass[letterpaper]{article}
\usepackage{aaai}
\usepackage{times}
\usepackage{helvet}
\usepackage{courier}
\usepackage{CJK}
\usepackage{algorithm}
\usepackage{algorithmic}
\usepackage{amsmath}
\usepackage{graphicx}
\usepackage{epstopdf}
\usepackage{subfigure}
\usepackage{float}
\usepackage{array}
\usepackage{titling}
\usepackage{booktabs}
\usepackage{multirow}
\usepackage{amssymb}
\usepackage{enumerate}
\usepackage{mathrsfs}
\usepackage{amsthm}
\usepackage{latexsym,bm}
\usepackage{times}
\usepackage{verbatim}
\begin{document}
%
\title{Understanding and Monitoring Human Trafficking via Social Sensors: A Sociological Approach}
\date{}
\author{Yang Yang\\
  \text{yangyangfuture@gmail.com}\\
  \text{Beihang University, China}
  \and 
   Xia Hu\\
    \text{hu@cse.tamu.edu}\\
    \text{Texas AM University, United States}
  \and 
   Haoyan Liu\\
    \text{haoyan.liu@buaa.edu.cn}\\
    \text{Beihang University, China}
    \and
     Zhoujun Li\\
    \text{lizj@buaa.edu.cn}\\
    \text{Beihang University, China}
    \and
     Philip S. Yu\\
    \text{psyu@uic.edu}\\
    \text{University of Illinois at Chicago, United States}
}
%

\maketitle
\begin{abstract}
Human trafficking is a serious social problem, and it is challenging mainly because of its difficulty in collecting and organizing related information. With the increasing popularity of social media platforms, it provides us a novel channel to tackle the problem of human trafficking through detecting and analyzing a large amount of human trafficking related information. Existing supervised learning methods cannot be directly applied to this problem due to the distinct characteristics of the social media data. First, the short, noisy, and unstructured textual information makes traditional  learning algorithms less effective in detecting human trafficking related tweets.
Second, complex social interactions lead to a high-dimensional feature space and thus present great computational challenges. In the meanwhile, social sciences theories such as homophily have been well established and achieved success in various social media mining applications. Motivated by the sociological findings, in this paper, we propose to investigate whether the \emph{\underline{N}etwork \underline{S}tructure \underline{I}nformation} (NSI) could be potentially helpful for the human trafficking problem.
In particular, a novel mathematical optimization framework is proposed to integrate the network structure into content modeling.
Experimental results on a real-world dataset demonstrate the effectiveness of our proposed framework in detecting human trafficking related information.


\end{abstract}
\section{Introduction}
Human trafficking \cite{mcgill2003human} is the trade in humans, and it is a crime because of the violation of the victims' rights of movement through coercion. For example, it includes the forced labor, extraction of organs or tissues and forced marriage, etc. It is also the fastest growing crime all over the world and one of the largest sources of income for organized crime. For each year, 600,000-800,000 adults and children are trafficked across international borders \cite{andrijasevic2007beautiful}. In addition, human trafficking not only happens in many developing countries but also in many developed countries, such as the USA and European countries.
Though many International-Governmental Organizations (IGO) and Non-Governmental Organizations (NGO) spent a lot of time and efforts to tackle this problem, human trafficking is still very challenging because of the following reasons.
First, many organizations lack sufficient and timely data. The human trafficking related data posted by these organizations is mainly manually collected from various sources, such as calls, emails and web applications. Also, while many organizations are collecting data by themselves, it is hard to share data between each other in a timely manner. Second, the data collected from multiple resources are unstructured and heterogeneous, and it presents great challenges to existing computational methods.
Thus we propose to examine this problem from a novel perspective in our study.

The increasing popularity of social media platforms provides a great opportunity to develop new approaches to help address human trafficking problem from the data perspective. The social networks have the data from both the victims of the human trafficking and their family members. For example, 1) some people are looking for their missing daughters through posting information online. 2) many victims of the human trafficking grow up, and they post the tweet on the social network to find their parents. There are also many children beggars, and they are the victims of human trafficking. The social network users upload information of the children beggars on social networks. Hence, social networks bridge the gap from both sides. It's meaningful to collect the tweets first and then match them to help people find their family members. However, useful information could be easily buried in the extremely large amount of social media posts. This motivates our study of identifying human trafficking related information on social networks, which could enable a number of related applications in the future.
%

Identification of human trafficking related information can be simply modeled as a classification problem based on the content information or network interactions. However, existing supervised learning methods cannot be directly applied because of the following reasons.
First, texts in social media are very short. Twitter allows users to post tweets with no more than 140 characters. Because of the short length, the performance of existing learning methods depending on similarity measurement between texts will be significantly affected.
Second, textual features are unstructured in social media. Many short texts are concise with many misspellings. Users describe the same thing in different and non-standard ways. For instance, some social network users use terms like ``gotta'', ``luv'' and ``wot''.
Third, it consists of complicated users interactions. In addition to content information, the interactions among users are essential but difficult to be used for our study. Thus it is challenging for existing learning methods to accurately identify the human trafficking information in social media data, which are short, unstructured and containing complex user interactions.

\begin{figure}
\centering
   \includegraphics[width=0.4\textwidth,angle=0]{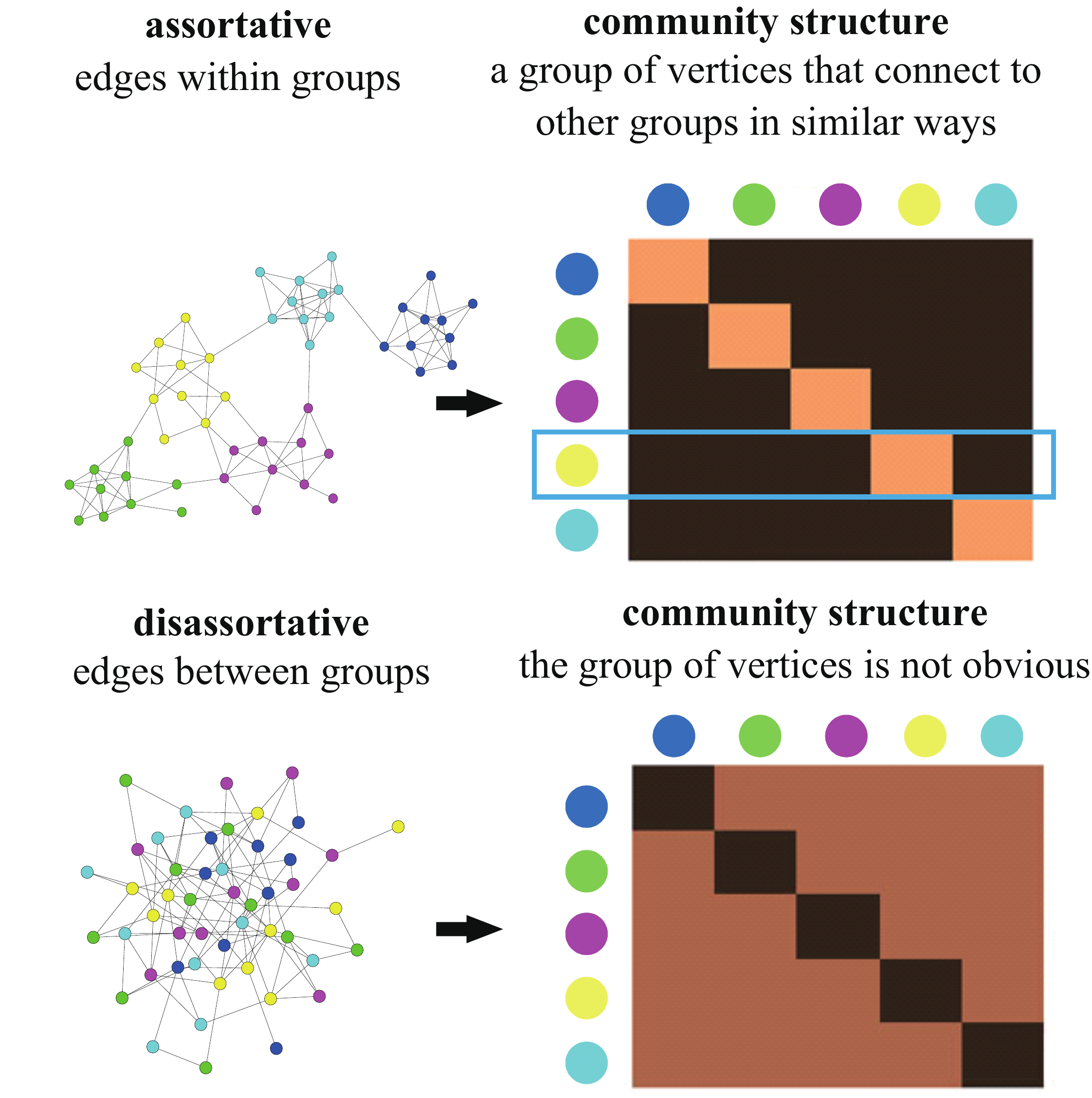}
  \caption{Assortative network and the corresponding community structure. Vertices in a community may have similar labels, and vertices that connect with each other probably have the same labels}
  \label{fig:ass}
\end{figure}

In the meanwhile, intensive efforts have been made to study the nontrivial properties of social networks \cite{newman2010networks,zhou2005learning}. It has been well established that, on social networks, users and features are correlated with each other and no longer independent. Users and tweets always form a complex assortative network with strong community structure. Based on the network theory, ``homophily", i.e. assortativity concept \cite{zhou2005learning,chung2005laplacians}, defines the extent for a network's vertices to attach to others that are similar in some way. As shown in Figure \ref{fig:ass}, the vertices in a community may have similar labels, and vertices that connect with each other probably have the same labels, which are helpful for vertices classification. Motivated by the above studies, we propose to investigate how assortativity could help identify human trafficking related information.

In this paper, we study the problem of identifying and understanding human trafficking on social media. Essentially, through our study, we want to answer the following questions.
1) How to define the problem of human trafficking text identification?
2) How to extract and select the content/network structure information?
3) How to handle the curse of dimensionality of text and structure features?
4) How to integrate network structure and content information in a model?
By answering the above questions, the main contributions of this paper can be summarized as follows:
\begin{itemize}
  \item Formally define the problem of human trafficking text identification with network structure and content features.
  \item Employ sparse learning methods to automatically select features and learn a simple model.

  \item Propose a unified model to effectively integrate network structure and content information, and a novel optimization framework is introduced.

  \item Evaluate the proposed NSI model on social network data and demonstrate the effectiveness of NSI.

\end{itemize}

\vspace{-3mm}
\section{Problem Description}
The notations are introduced in this section, and then we define the problem we study.
$||\textbf{A}||_{F}$ denotes the Frobenius norm of the matrix $\textbf{A}\in \mathbb{R}^{m\times n}$, where $n$ is the number of tweets and $m$ is the number of features, and $||\textbf{A}||_{F}$ is defined as $|| \textbf{A} ||_{F}= \sqrt{\sum_{i=1}^{m}\sum_{j=1}^{n}|a_{ij}|^{2}} =\sqrt{trace(\textbf{A}^{T}\textbf{A}))}$.
$||\textbf{W}||_{1}$ is the $\ell_{1}$ norm of the weights vector $\textbf{W}\in \mathbb{R}^{m\times c}$, where $c$ is the number of categories, and the $||\textbf{W}||_{2,1}=\sum_{i=1}^{m}{\sqrt{\sum_{j=1}^{c}{W^{2}_{i,j}}}}$ is the $\ell_{2,1}$ norm of the weight matrix $\textbf{W}$.
The trace of a matrix $\textbf{B}$ is $tr(\textbf{B})$, and the transpose of matrix $\mathbf{B}$ denotes as $\textbf{B}^{T}$.

\begin{figure}[htbp]
  \centering

    \includegraphics[width=0.4\textwidth,angle=0]{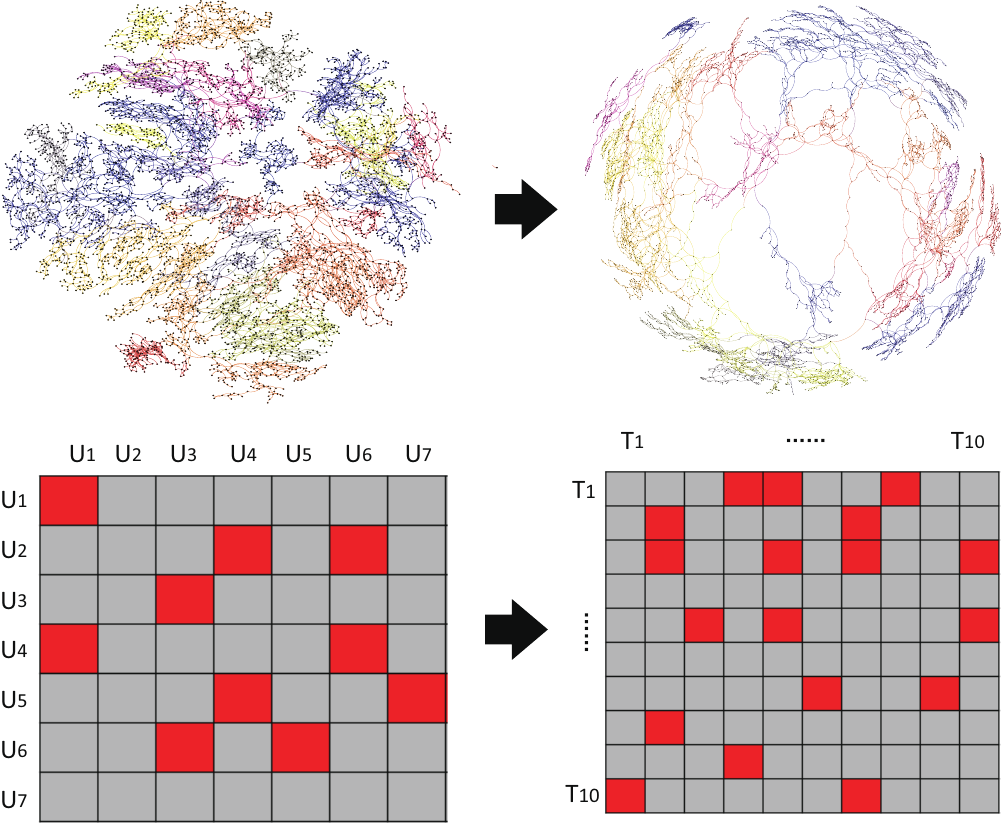}

  \caption{Reformulation of the user behavior network. The original network is represented as a user-user matrix. The converted network is represented as a tweet-tweet matrix. We convert the original network into a new network.}
  \label{fig:matrix}
  \vspace{-2mm}
\end{figure}

Graph $G=(B,F)$ denotes the reply and retweet relationship among users, in which vertices $u$ and $v$ represent users in social media, while the edges in $F$ represent the reply or retweet relationships among the users. To classify the tweets into two classes, i.e. the tweets are about human trafficking or not, the task is to classify the edges in graphs into two classes (positive or negative) --- an edge classification problem. However, the methods that classify edges into two classes are not common. Hence, we convert node adjacency matrix of $G=(B,F)$ to the edge adjacency matrix of $H=(U,V)$, as shown in Figure \ref{fig:matrix}. The main idea of the conversion algorithm is that if two edges have the same start or end node, there is a link between them.
\vspace{-2mm}
\\\emph{Here we formally define the problem we study as follows.
Given a set of microblogging tweets $U$ with network information $H=(U,V)$, content information $\mathbf{X}\in \mathbb{R}^{m\times n}$, and identity label information $\mathbf{Y}\in \mathbb{R}^{n\times c}$, our aim is to learn a classifier $\mathbf{W}\in \mathbb{R}^{m\times c}$ to automatically identify human trafficking related tweets from unknown tweets.}
\vspace{-3mm}
\section{A Sociological Approach---NSI}
In this section, we firstly model the social network structure information for tweet classification. Then we discuss the methods in modeling the content information. At last, we integrate the social network structure information and the content information into a unified model and propose an optimization algorithm to solve the human trafficking tweets recognition problem.
\vspace{-2mm}
\subsection{Modeling Network Information}
Network information plays an important role in solving practical problems, as it contains much useful information which cannot be mined in pure content information. In addition to pure content information, many studies have employed network information in solving real-world problems: sentiment analysis, influential users \cite{rabade2014survey}, recommendation \cite{tang2013social} and topic detection \cite{Chenetal12}. The concepts ``homophily" and ``community structure" in social sciences indicate that the vertices in each community and the vertices connected with each other probably have similar labels. Motivated by these theories, we employ homophily and community structure to help identify human trafficking tweets. 

Existing work \cite{platt1999large,hansen1996classification} has been done to classify the vertices in networks into two classes. These methods are all on directed networks. Hence, we see the edge in our undirected network as a bidirectional edge. Similarly, the vertice $u$'s in-degree is defined as $d^{in}_{u}=\sum_{[v,u]}{H(v,u)}$, and the vertice $u$'s out-degree is defined as $d^{out}_{u}=\sum_{[u,v]}{H(u,v)}$. $\mathbf{P}$ is defined as the transition probability matrix of random walk in a graph with $P(u,v)=H(u,v)/d^{out}_{u}$. The stationary distribution $\pi$ of the random walk satisfies: $\sum_{u\in V}{\pi(u)}=1,~ \pi(v)=\sum_{[u,v]}{\pi(u)P(u,v)}.$
The network information is used to smooth the unified model. 
The classification problem can be formulated as minimizing
\begin{align}
    R(f):=\frac{1}{2}\sum_{[u,v]\in E}{\pi(u)P(u,v)||\hat{Y_{u}}-\hat{Y_{v}}||^{2}},
    \label{equ:rs}
\end{align}
where $\hat{Y_{u}}=\frac{f(u)}{\sqrt{\pi{(u)}}}$ is the predicted label of user $u$, and $\hat{Y_{v}}=\frac{f(v)}{\sqrt{\pi{(v)}}}$ is the predicted label of user $v$. $\mathcal{H}(V)$ is the function space, and $f\in \mathcal{H}(V)$ is the classification function, which assigns a label sign $f(v)$ to each vertex $v\in V$. If two tweets $u$ and $v$ are close to each other and have different predicted labels, the above loss function will have a penalty. For solving the Equation (\ref{equ:rs}), we introduce an operator $\Theta: \mathcal{H}(V)\rightarrow \mathcal{H}(V)$.
\begin{align}
\begin{split}
(\Theta f)(v) = &\frac{1}{2}\bigg(\sum_{u\rightarrow v}{\frac{\pi(u)P(u,v)f(u)}{\sqrt{\pi(u)\pi(v)}}}\\
&+\sum_{u\leftarrow v}{\frac{\pi(v)P(v,u)f(u)}{\sqrt{\pi(v)\pi(u)}}}\bigg).
\end{split}
\end{align}
It has been shown \cite{hansen1996classification} that the objective function can be interpreted as $R(f) = tr(\hat{Y}\mathcal{L}\hat{Y}^{T})$,
where the $\mathcal{L}=I-\Theta$.
\begin{align}
    \Theta = \frac{\mathbf{\Pi}^{1/2}\mathbf{P}\mathbf{\Pi}^{-1/2}+\mathbf{\Pi}^{-1/2}\mathbf{P}^{T}\mathbf{\Pi}^{1/2}}{2},
\end{align}
where $\mathbf{\Pi}$ is a diagonal matrix with entries $\mathbf{\Pi}(v,v)=\pi(v)$. $\pi$ denotes the eigenvector of the transition probability $\mathbf{P}$, and $\mathbf{P}^{T}$ is the transpose of $\mathbf{P}$.
If the original network is an undirected network, the $\mathcal{L}$ is reduced to $\mathbf{D}-\mathbf{A}$. $\mathcal{L}$ is symmetric and positive-semidefinite. $\mathbf{D}$ is the degree matrix and $\mathbf{A}$ is the adjacency matrix of the graph.

\subsection{Modeling Content Information}
The most important task is to distinguish the human trafficking tweets from other social media posts. The text information \cite{hu2012text} is necessary for the task. 

One of the most widely used methods is the Least Squares \cite{lawson1974solving}, which is an efficient and interpretable model. The classifiers can be learned by solving the following equation:
\begin{eqnarray}
    \underset{\mathbf{W}}{min} ~\frac{1}{2} ||\mathbf{X}^{T}\mathbf{W}-\mathbf{Y}||_{F}^{2},
\end{eqnarray}
where $\mathbf{X}$ is the content feature matrix of the training data, and $\mathbf{Y}$ is the label matrix. This formulation is to minimize the learning error between the predicted value $\mathbf{\hat{Y}} = \mathbf{X}^{T}\mathbf{W}$ and the true
value $\mathbf{Y}$ in the training data.

However, high-dimensional feature space makes the computational task extremely difficult. As we know, sparse learning method has shown its effectiveness in many real-world applications such as \cite{pu2006short} and \cite{tibshirani1996regression} to handle the high-dimensional feature space. Hence, we propose to make use of the sparse learning for selecting the most effective features. Sparse learning methods \cite{liu2009slep,chen2009accelerated} are widely used in many areas, such as the sparse latent semantic analysis and image retrieval. The sparse learning methods can generate a more efficient and interpretable model. A widely used regularized version of least squares is the lasso (least absolute shrinkage and selection operator) \cite{tibshirani1996regression}. Hence, we can get a text classifier through solving the $\ell_{1}$-norm penalization on least squares:
\begin{eqnarray}
    \underset{\mathbf{W}}{min} ~\frac{1}{2}||\mathbf{X}^{T}\mathbf{W}-\mathbf{Y}||_{F}^{2}+\lambda_{1}||\mathbf{W}||_{1}.
\end{eqnarray}


In many real world applications, the features are not independent. In our task, the features are related with each other according to the same POS tagging. These relations among features may have a positive effect on the classification. Hence, to analyze the complex features, we introduce the regularization $\ell_{2,1}$-norm to the model, which is considered as one of the most popular ones due to its effectiveness, robustness, and efficiency. It is defined as $||\mathbf{W}||_{2,1}=\sum_{i=1}^{m}||w_{g_{i}}||_{2}$, where $g_{i}$ is a group of features. Another appealing feature of the $\ell_{2,1}$-norm regularization is that it encourages multiple predictors to share similar sparsity patterns \cite{liu2010fast}. And the $\ell_{2,1}$-norm regularization can automatically select variables, do continuous shrinkage and select groups of correlated variables \cite{friedman2010note,eksioglu2014group,chen2010regularized,bengio2009group}. Hence, we get a more stable model by solving the equation:
\begin{eqnarray}
    \underset{\mathbf{W}}{min}~\frac{1}{2}||\mathbf{X}^{T}\mathbf{W}-\mathbf{Y}||_{F}^{2}+\lambda_{1}||\mathbf{W}||_{1}+\frac{\lambda_{2}}{2}||\mathbf{W}||_{2,1},
\end{eqnarray}
where $\lambda_{1}$ and $\lambda_{2}$ are positive regularization parameters to control the sparsity and robustness of the learned model.
\vspace{-2mm}
\subsection{The Objective Function}
Traditional text classification methods \cite{Liu2011Partially,Oren2005A} intend to add new features or propose effective classifiers to successfully solve the problem.
On the one hand, the dimension of the text feature is always high. Traditional methods are not able to handle high dimension features. These methods have to select features first, and then learn a model to classify the texts.
Sparse learning method, which can automatically select features and learn a model, is a good choice to solve the problem.
On the other hand, one assumption of many traditional methods is that the features of texts are independent. However, some features of the texts are correlated with others.
It is of great importance if we can consider networked or grouped features in the learning task. Hence, we employ the sparse group lasso to analyze the relationships among different features.

Furthermore, network structure information plays an important role in identifying the human trafficking tweets. The assortativity and community structure are used to formulate the behaviors among users. The behavior information contains much useful information that text information doesn't have. Hence, we further integrate the two kinds of features.

We propose to consider both network and content information in a unified model.
By considering both network and content information, the human trafficking tweets recognition problem can be formulated as the optimization problem:
\begin{align}
    \underset{\mathbf{W}}{min}~\quad&O(\mathbf{W})=\frac{1}{2}||\mathbf{X}^{T}\mathbf{W}-\mathbf{Y}||_{F}^{2}+\lambda_{1}||\mathbf{W}||_{1}+\frac{\lambda_{2}}{2}||\mathbf{W}||_{2,1}\notag\\
    &+\frac{\lambda_{s}}{2}tr(\mathbf{W}^{T}\mathbf{X}\mathbf{L}\mathbf{X}^{T}\mathbf{W}).
    \label{equ:objective}
\end{align}
\emph{The $\mathbf{W}$ can be derived from solving the above equation. Then, we can use the following equation to predict the label of an unknown tweet}:
\begin{align}
\underset{i\in \{0, 1\}}{arg~~max}~~x^{T}w_{i}.
\end{align}
\subsection{An Optimization Algorithm}
The optimization problem in Equation (\ref{equ:objective}) is convex and nonsmooth. We intend to reformulate the non-smooth optimization problem, and get an equivalent smooth convex optimization problem, according to the main idea of \cite{liu2010fast,nesterov2004introductory}.\\
\textbf{}By converting the $\ell_{1}$-norm as the constrained condition, we can reformulate the Equation (\ref{equ:objective}) as a constrained nonsmooth convex optimization problem:
\begin{align}
\begin{split}
    \underset{\mathbf{W}\in \mathbf{Z}}{min}\quad O(\mathbf{W})=&||\mathbf{X}^{T}\mathbf{W}-\mathbf{Y}||_{F}^{2}+\lambda_{2}||\mathbf{W}||_{2,1}\notag\\
    &+\lambda_{s}tr(\mathbf{W}^{T}\mathbf{X}\mathbf{L}\mathbf{X}^{T}\mathbf{W}),
\end{split}
\end{align}
where
\begin{align}
\begin{split}
  Z=\{\mathbf{W}|~||\mathbf{W}||_{1}\leqslant z\}.
\end{split}
\label{equ:final}
\end{align}
The $||\mathbf{W}||_{1}$ defines a closed and convex set $Z$. The objective function $O(\mathbf{W})$ is a convex but not differentiable, as the function $||\mathbf{W}||_{2,1}$ is differentiable everywhere, except when $||w_{g_{i}}||_{2}=0$ or equivalently when $\mathbf{W}$ is the all zero vector $0\in \mathbb{R}^{n}$ \cite{liu2010fast,eksioglu2014group}. At the point $w_{g_{i}}=0$, the subdifferential includes the all zero vector as a legitimate subgradient, that is, $0\in{\partial ||w_{g_{i}}||_{2}}$ when $w_{g_{i}}=0$.
Motivated by the work \cite{nie2010efficient,chen2010regularized}, we propose an iterative algorithm. Through denoting $\mathbf{D}$ as a diagonal matrix $D_{ii}=\frac{1}{2||w_{g_{i}}||_{2}}$, Equation (\ref{equ:final}) can be formulated as a convex and smooth optimization:
\begin{align}
  \underset{\mathbf{W}\in Z}{min}\quad O(\mathbf{W})=&||\mathbf{X}^{T}\mathbf{W}-\mathbf{Y}||_{F}^{2}+\lambda_{2}tr(\mathbf{W}^{T}\mathbf{D}\mathbf{W})\notag\\
    &+\lambda_{s}tr(\mathbf{W}^{T}\mathbf{X}\mathbf{L}\mathbf{X}^{T}\mathbf{W}).
\end{align}
The derivative of $O(\mathbf{W})$ is as follows.
\begin{align}
\frac{\partial O(\mathbf{W})}{\partial \mathbf{W}}=\mathbf{X}\mathbf{X}^{T}\mathbf{W}-\mathbf{X}\mathbf{Y}+\lambda_{2}\mathbf{D}\mathbf{W}+\lambda_{s}\mathbf{X}\mathbf{L}\mathbf{X}^{T}\mathbf{W}.
\end{align}
\vspace{-2mm}
Let the derivative of $O(\mathbf{W})$ be equal to zero.
\begin{align}
&\mathbf{X}\mathbf{X}^{T}\mathbf{W}-\mathbf{X}\mathbf{Y}+\lambda_{2}\mathbf{D}\mathbf{W}+\lambda_{s}\mathbf{X}\mathbf{L}\mathbf{X}^{T}\mathbf{W}=0\notag\\
&\Leftrightarrow (\mathbf{X}\mathbf{X}^{T}+\lambda_{2}\mathbf{D}+\lambda_{s}\mathbf{X}\mathbf{L}\mathbf{X}^{T})\mathbf{W}=\mathbf{X}\mathbf{Y}\notag\\
&\Leftrightarrow \mathbf{W}=(\mathbf{X}\mathbf{X}^{T}+\lambda_{2}\mathbf{D}+\lambda_{s}\mathbf{X}\mathbf{L}\mathbf{X}^{T})^{-1}\mathbf{X}\mathbf{Y}.
\end{align}

The solution of $\mathbf{W}$ is correlated with the input of $\mathbf{D}$ which is also related to $\mathbf{W}$. Therefore, the $\mathbf{W}$ cannot be obtained directly. Instead we propose an algorithm to optimize the parameters in Algorithm \ref{alg_lirnn}, according to \cite{nesterov2004introductory,liu2009multi}.
\begin{algorithm}[ht]
\caption{\small Identifying the Human Traf{f}icking Information Problem}
\textbf{Input:} $\mathbf{X},\mathbf{Y},\mathbf{L},\mathbf{W_{0}},\lambda_{1},\lambda_{2},\lambda_{s}$. \\
\textbf{Output:} $\mathbf{W}$.
\begin{enumerate}[1:]

\item $\tau=1$; Initialize the \textbf{W} as an identity matrix;
\item \textbf{repeat}
\item      ~~~~~~Update the diagonal matrix \textbf{D} as
\item
~~~~~~$\mathbf{D}_{\tau+1}=$$\begin{bmatrix}
\frac{1}{2||{w^{\tau}_{g_{i}}}||_{2}} &  & \\
 &.....  & \\
 &  & \frac{1}{2||{w^{\tau}_{g_{m}}}||_{2}}
\end{bmatrix}$\\
\item ~~~~~~Update the metric \textbf{W}:\\
  $~~~~~ ~~~~~ \mathbf{W}_{\tau+1}=(\mathbf{X}\mathbf{X}^{T}+\lambda_{2}\mathbf{D_{\tau}}+\lambda_{s}\mathbf{X}\mathbf{L}\mathbf{X}^{T})^{-1}\mathbf{X}\mathbf{Y}$;

\item \textbf{until} convergence

\item $\mathbf{W}=\mathbf{W}_{\tau+1}$
\item $PSD$ projection: $\mathbf{W}=PSD(\mathbf{W})$

\end{enumerate}
\label{alg_lirnn}
\vspace{-2mm}
\end{algorithm}
The basic idea of the proposed algorithm is to reformulate the nonsmooth optimization problem as an equivalent smooth convex optimization problem. In the algorithm, we use the $\mathbf{D}_{\tau}$ to continually update the $\mathbf{W}_{\tau}$ from lines 1 to 6. The positive semi-definite (PSD) constraint of $\mathbf{W}$ is not necessary for each iterative step. We perform the PSD projection for the final weights matrix $\mathbf{W}$ for efficiency at lines 7-8. The $\lambda_{1}$ is set in projection function.\\
\vspace{-6mm}
\subsection{Time complexity analysis}
It takes $O(n^{2})$ to calculate the laplacian matrix $\mathbf{L}$ before the iterative procedure, where $n$ is the number of samples in the Laplacian Matrix. The procedure of updating $\mathbf{D}$ and the inverse of the matrix $\mathbf{X}\mathbf{X}^{T}+\mathbf{D}+\lambda \mathbf{X}\mathbf{L}\mathbf{X}^{T}$ need $O(m^{3})$. 
The item $\mathbf{D}$ which is a diagonal matrix makes the inverse of $\mathbf{X}\mathbf{X}^{T}+\mathbf{D}+\lambda \mathbf{X}\mathbf{L}\mathbf{X}^{T}$ more stable. We need $O(m^{2})$ to obtain W. Hence, each iterative step costs $O(m^{3}+m^{2})$. Suppose the optimization algorithm takes $T$ iterations, the overall time complexity is $O(Tm^{3}+Tm^{2})$. 
\vspace{-4mm}
\section{Experiments}
We intend to evaluate the effectiveness of the proposed method in this paper and analyze the effectiveness of the network structure and content information. The experiments in this section focus on solving the following questions:
\begin{enumerate}
\item  How effective is the proposed method compared with the baseline methods?
\item  What are the effects of the network structure and content information?
\end{enumerate}

We begin by introducing data set and experimental setup, and then
compare the performance of different traditional machine learning methods. Finally, we study the effects of network and content features on the proposed method.
\vspace{-3mm}
\subsection{Data set}
The real-world weibo data set is firstly crawled from September 2014 to October 2014. Weibo is a Chinese microblogging website that is akin to the Twitter.
We generally sample 14151 tweets which contain the keywords ``Human trafficking" and ``missing people". Then, 5 students annotate 14151 tweets with 1,404 positive samples. All the tweets are marked as positive and negative according to whether the weibo is related to human trafficking.
Each tweet is retweeted or replied by 16.8 times on average. The retweet/reply frequencies follow the power law distribution, which indicates that few of the tweets draw much attention
, and most of the tweets are neglected.

We investigate these tweets and propose general and domain specific features：
\begin{itemize}
\item Word based features: unigrams, unigrams+bigrams, and Pos colored unigrams+bigrams. Pos tagging is done by the Jieba package. When the corpus is large, the dimensions of the unigrams and unigram+bigrams features are too high for a PC to handle. Hence, we pick up the Pos colored unigrams+bigrams feature.
\item Tag based features: Most of the human trafficking tweets have tags. Having a tag in the tweet may promote more users to reach the information.
\item Morphological feature: These include each feature for the frequencies of 
    \begin{itemize}
    \item the number in the sentence
    \item the question mark in the sentence
    \item the exclamation mark in the sentence
    \item the quantifiers in the sentence
    \end{itemize}

\item NER features Most of the human trafficking related tweets contain a name, location, organization and time.
\item Tweet features: the length of tweets

\end{itemize}


\vspace{-4mm}

\subsection{Performance Evaluation}
Precision, recall and F1-measure are used as the performance metrics. The F1 measure is the harmonic mean of precision and recall. 
To answer the first question, we compare our proposed method with the baseline methods: SVM \cite{suykens1999least}, Logistic Regression (LR) \cite{hosmer2004applied}, Gaussian Naive Bayes (GNB) \cite{john1995estimating}, SGD \cite{xiao2009dual}, Decision Tree (DT) \cite{friedl1997decision} and Random Forests (RF) \cite{liaw2002classification}. All the methods utilize both social network structure and content information. The combined feature is the linear combination of the social network structure feature and content feature.

\begin{figure}[htbp]
\centering
  \includegraphics[width=0.45\textwidth,angle=0]{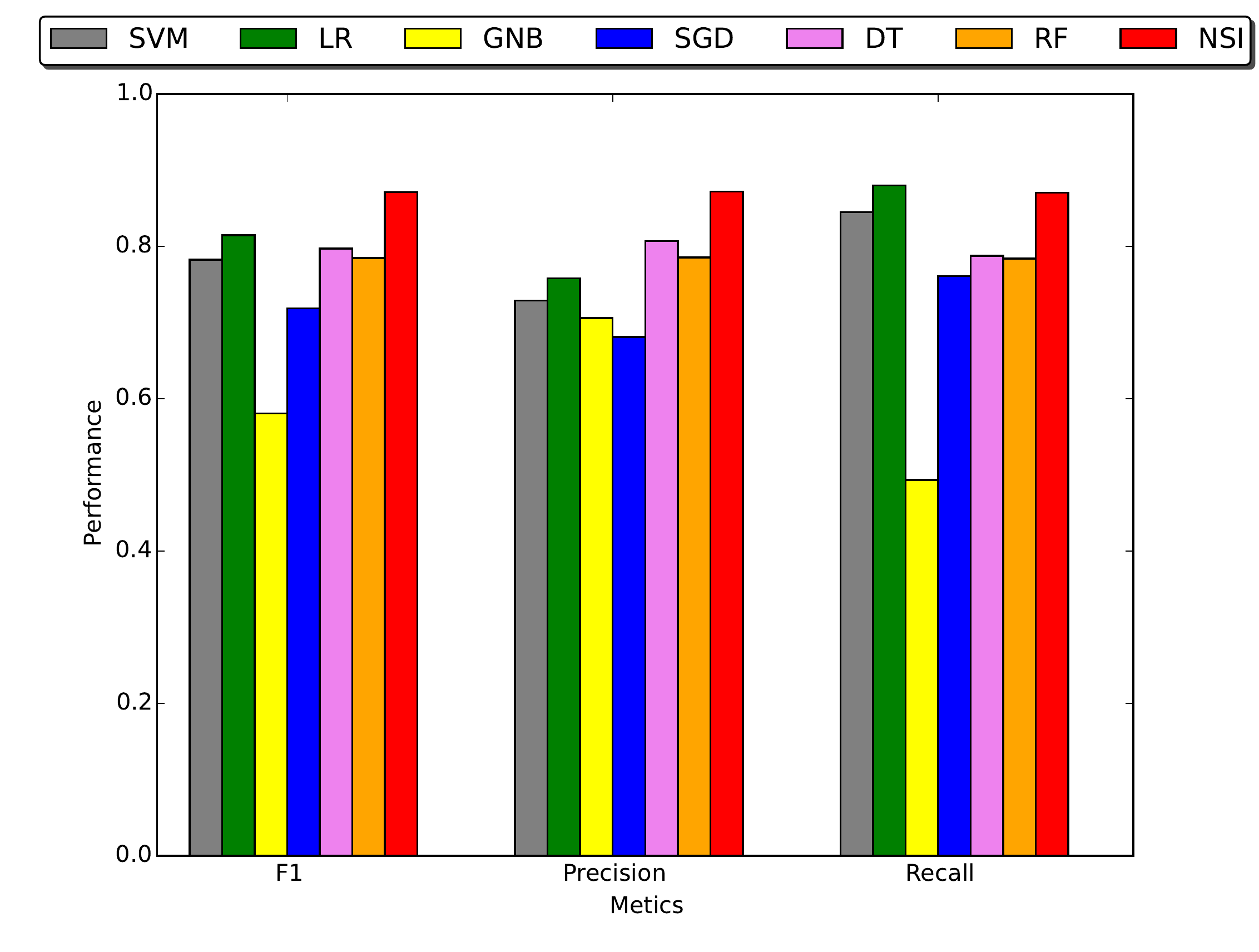}
  \vspace{-2mm}
 \caption{The performance of baseline methods and NSI model. We classify the tweets with social network and content information.}
 \label{fig:yy} 
 \vspace{-4mm}
\end{figure}

(1) According to the results in Table \ref{tab:contentandnetwork} and Figure \ref{fig:yy}, we can draw a conclusion that our model NSI outperforms other methods in precision and F1-measure. Two-sample one tail t-tests are applied to compare NSI with other methods in Section \uppercase\expandafter{\romannumeral4}, which indicates the effectiveness of our model.

(2) As shown in Figure \ref{fig:yy}, the NSI model, decision tree and random forests have relatively balanced precision and recall. 
Most of the baseline methods have a high recall with small precision.
The performance of SGD is not as good as that of SVM. The reason is that the poor performance of L2 regularization is linked to rotational invariance.
\begin{table}[htbp]
  \centering
  \caption{Classify tweets with network and content features}
    \begin{tabular}{lccc}
    \toprule
    Methods & \multicolumn{1}{c}{F1} & \multicolumn{1}{c}{Precision} & \multicolumn{1}{c}{Recall} \\
            \midrule
    SVM   & 0.783 & 0.729 & 0.845 \\
    LR    & 0.815 & 0.758 & 0.880 \\
    GNB   & 0.581 & 0.706 & 0.493 \\
    SGD   & 0.719 & 0.681 & 0.761 \\
    DT    & 0.797 & 0.807 & 0.788 \\
    RF    & 0.785 & 0.786 & 0.784 \\
    NSI   & 0.871 & 0.872 & 0.870 \\
                \bottomrule
    \end{tabular}%
  \label{tab:contentandnetwork}%
\end{table}%

The performance of all the methods cannot solve the classification problem perfectly. The reason is as follows:\\
1) Though we provide some instructions for every annotator, some tweets are so ambiguous that they cannot distinguish the class of the tweet. \\
2) Some tweets search for an unfamiliar charming boy/girl that the user met by accident in the real world. It's a ``searching'' people tweet. And the feature of this kind of tweet is similar to the human trafficking related tweets.
\vspace{-4mm}

\subsection{Case study}
Table \ref{tab:case-study} shows the examples of human trafficking tweets. The first column is the label of the tweet. The second column is the tweet content. The name of missing people and HTTP link are replaced with hashtag \#Name\# and \#http\#, respectively. The 1st tweet in Table \ref{tab:case-study} is a typical human trafficking tweet. It contains detailed information of the missing people: name, age, height, and clothing. The 3rd tweet is also a human trafficking tweet posted by the victim of the human trafficking. We intend to identify these tweets first and then match them. Almost all machine learning methods could correctly label the 1st, 3rd, 4th and 5th tweet, while traditional machine learning methods label the 2nd tweet as a negative sample. The reason is that this tweet doesn't contain any detailed information. However, our model could correctly classify the tweet, as this tweet is retweeted by many users who ever retweeted many human trafficking tweets. These retweet/reply behaviors are incorporated into the Laplacian matrix in our model. By introducing the matrix, the model smooths the weight of $W$. It leads to increase the precision of the model but decrease the recall to some extent. That's the reason why our model gets a good performance with a relative balanced precision and recall.

\begin{table}[]
\centering
\caption{A case study of human trafficking tweets}
\label{tab:case-study}
\resizebox{\columnwidth}{!}{%
\begin{tabular}{|c|l|}
\hline
Label &  \multicolumn{1}{c|}{Tweet}                                                                                                                                                                                                                                                       \\ \hline
1     & \begin{tabular}[c]{@{}l@{}}\#NAME\# is a missing people, 21 yrs old. She was last seen on Sep \\19 in the downtown. 5'6" in height. Last seen wearing a green top\\with a red striped Adidas jacket over top and carrying a white purse.\end{tabular} \\ \hline
1     & \#Name\# from Hubei was last seen at 10am \#http\#                                                                                                                                                                                                               \\ \hline
1 & \begin{tabular} [c]{@{}l@{}}I was trafficked from ShanXi to Hebei in 1983. Type B blood.\\ Height 164cm. Weight 70kg. I was born in Datong or Taiyuan.\\ The trafficker is from Hebei. I want to find my parents.\end{tabular}
\\ \hline
0     & \begin{tabular}[c]{@{}l@{}}Baby dead, 3 people missing, 3 others injured after \\ extra-alarm fire in South.\end{tabular}                                                                                                                                \\ \hline
0     & Missing people that don't miss me.                                                                                                                                                                                                                              
\\ \hline
\end{tabular}
}
\end{table}

\vspace{-3mm}
\subsection{Effects of Network and Content Information}
In this subsection, we study the performance of network structure and content information, and compare our proposed method with the following two groups methods: 1) content feature: the traditional methods are employed for classifying human trafficking tweets based on content information only. 2) network feature: the traditional methods are applied on the network information.
%


(1) Considering the content feature only, most of the methods cannot balance the precision and recall, as shown in Table \ref{tab:contentornetwork}. The precision of SGD reaches 97\%, while its recall is 34.8\%.
SVM, logistic regression, and Gaussian have the higher recall, while other methods can achieve high precision.
Our NSI model which archives relatively balanced precision and recall has the highest F1-value.

(2) Considering the network feature only, almost all methods fail to classify the tweets into the right category. The precision of GNB is 79.3\%, which is the highest among other methods, while its recall is just 29.6\%. Different from GNB, the recall of the SVM is 93.8\%, however the precision of the SVM is only 56.9\%.

(3) In conclusion, considering the integrated features, the recall of all the methods rise except SVM, Gaussian naive Bayes, and SGD, while the precision of all the methods rises except SVM, logistic regression and SGD. All the methods have relatively balanced precisions and recalls. The reason is that the network structure  features successfully smooth the weight of the content feature. Hence, these features play an important role in predict the true label of a new tweet. 
\vspace{-4mm}

\begin{table}[htbp]
  \centering
  \caption{Classify tweets with network or content features}
\resizebox{\columnwidth}{!}{%
    \begin{tabular}{c|cccccc}
    \toprule
    \multirow{2}[4]{*}{Methods} & \multicolumn{3}{c}{Content feature} & \multicolumn{3}{c}{Network feature} \\
\cmidrule{2-7}          & F1    & Precision & Recall & F1    & Precision & Recall \\
    \midrule
    SVM   & 0.834 & 0.744 & 0.948 & 0.708 & 0.569 & 0.938 \\
    LR    & 0.823 & 0.780  & 0.871 & 0.694 & 0.571 & 0.885 \\
    GNB   & 0.755 & 0.705 & 0.812 & 0.431 & 0.793 & 0.296 \\
    SGD   & 0.512 & 0.970  & 0.348 & 0.512 & 0.538 & 0.489 \\
    DT    & 0.764 & 0.793 & 0.738 & 0.427 & 0.648 & 0.318 \\
    RF    & 0.769 & 0.769 & 0.770  & 0.331 & 0.348 & 0.315 \\
    NSI   & 0.855 & 0.856 & 0.854 & 0.409 & 0.695 & 0.290 \\
    \bottomrule
    \end{tabular}%
    }

  \label{tab:contentornetwork}%
\end{table}%

\vspace{-2mm}



%
%
%
%
%
\vspace{-4mm}
\subsection{Parameter Analysis}
There are three positive parameters involved in the experiments, including $\lambda_{1}$, $\lambda_{2}$, and $\lambda{s}$, as shown in Algorithm \ref{alg_lirnn}. $\lambda_{1}$ is to control the sparsity of the learned model, $\lambda_{2}$ is the parameter to control the group sparsity, and $\lambda{s}$ is to control the contribution of network information. As a common practice, all the parameters can be tuned via cross-validation with validation data. In the experiments, we empirically set $\lambda_{1} = 0.01$,$\lambda_{2} = 0.5$, and $\lambda_{s} = 0.4$ for general experiment purposes.
In this section, we will further explore the relation between the two parameters.
%
%
%

The performance of NSI model gets better when the $\lambda_{2}$ and $\lambda_{s}$ increase. The F1-value reaches the peak at $\lambda_{s}=0.4$ and $\lambda_{2}=0.5$. As the further increase of $\lambda_{2}$ and $\lambda_{s}$, the F1-value begins to drop. It indicates that the NSI model can achieve good performance when $\lambda_{s}\in[0.45,0.55]$ and $\lambda_{2}\in[0.5,0.6]$.
    \vspace{-6mm}
\section{Conclusion and Future Work}
The emergence of social networks provides a great opportunity to recognize and understand the human trafficking tweets. In this paper, we exploit the social network structure information to perform effective human trafficking tweets recognition. In particular, the proposed NSI models the network and content information in a unified way. An efficient algorithm is proposed to solve the non-smooth convex optimization problem. Experimental results on a real weibo data set indicate that our model NSI can effectively detect human trafficking tweets, and outperform alternative supervised learning methods.

There still exists open questions in tackling the problem of human trafficking. The extensions of this work are as follows. Other sparse learning methods can be introduced to analyze the complex structure of textual features. Many human trafficking tweets contain images. Hence, face/gender/age detection algorithms on the images are of great importance. It is meaningful to utilize the semantic analysis across social media sites to better understand the characteristics of the victims of the human trafficking with text features. How to utilize the distributed system to analyze the large data and solve the optimization problem is a promising direction.
\bibliographystyle{aaai}
\bibliography{formatting-instructions-latex}

\end{document}